\documentclass[12pt]{article}
\usepackage{epsfig}
\usepackage{graphicx}
\usepackage{color}
\def\Title#1{\begin{center} {\Large #1 } \end{center}}

\newenvironment{Abstract}{\begin{quotation}  }{\end{quotation}}
\newenvironment{Presented}{\begin{quotation} \begin{center} 
             PRESENTED AT\end{center}\bigskip 
      \begin{center}\begin{large}}{\end{large}\end{center} \end{quotation}}

\def\beq{\begin{equation}}
\def\eeq#1{\label{#1}\end{equation}}
\def\eeqn{\end{equation}}

\def\beqa{\begin{eqnarray}}
\def\eeqa#1{\label{#1}\end{eqnarray}}
\def\eeqan{\end{eqnarray}}

\let\bar=\overbar

\def\Dslash{\not{\hbox{\kern-4pt $D$}}}
\def\dslash{\not{\hbox{\kern-2pt $\del$}}}

\def\msb{{\bar{\ssstyle M \kern -1pt S}}}

\newcommand{\pippim}{\ensuremath{B^{0} \rightarrow \pi^{+} \pi^{-}}}

\newcommand{\rhorho}{\ensuremath{B \rightarrow \rho \rho}}

\newcommand{\rhozrhoz}{\ensuremath{B^{0} \rightarrow \rho^{0} \rho^{0}}}

\newcommand{\Ep}{\ensuremath{e^{+}}}
\newcommand{\Em}{\ensuremath{e^{-}}}
\newcommand{\pip}{\ensuremath{\pi^{+}}}
\newcommand{\piz}{\ensuremath{\pi^{0}}}
\newcommand{\pim}{\ensuremath{\pi^{-}}}
\newcommand{\aone}{\ensuremath{a_{1}(1260)}}
\newcommand{\Bp}{\ensuremath{B^{+}}}
\newcommand{\Bz}{\ensuremath{B^{0}}}
\newcommand{\Bzb}{\ensuremath{\bar{B}^{0}}}
\newcommand{\YFS}{\ensuremath{\Upsilon(4S)}}

\newcommand{\BBbar}{\ensuremath{B\bar{B}}}

\newcommand{\Mbc}{\ensuremath{M_{\rm bc}}}

\newcommand{\De}{\ensuremath{\Delta E}}
\newcommand{\Dt}{\ensuremath{\Delta t}}

\newcommand{\Acp}{\ensuremath{{A}_{CP}}}

\newcommand{\Scp}{\ensuremath{{S}_{CP}}}

\newcommand{\phitwo}{\ensuremath{\phi_{2}}}

\newcommand{\rz}{\ensuremath{\rho^{0}}}

\begin{document}
%\begin{titlepage}
%\pubblock
\Title{Results for $\phitwo$ from Belle}
\bigskip\bigskip

\begin{raggedright}  
{\it Pit Vanhoefer\index{Vanhoefer, P.}\\
Max-Plank-Institut f\"ur Physik\\
 F\"oringer Ring 6\\
  M\"unchen 80805 GERMANY}
\bigskip\bigskip
\end{raggedright}

%\vfill
%\Title{Results for $\phione$ and $\phitwo$ from Belle}
%\vfill
%\Author{ P.~Vanhoefer}
%\Address{\mpi}
%\vfill

\begin{Abstract}
We present a summary of measurements sensitive to the CKM angle \phitwo, performed by the Belle experiment using the final data sample of $772 \times 10^{6}$ \BBbar\ pairs produced at the \YFS\ resonance at the KEK asymmetric \Ep\Em\ collider. We discuss $CP$ asymmetries from the decay $B \rightarrow \pi^{+} \pi^{-}$ and briefly mention a preliminary measurement of the branching fraction of $B\to\piz\piz$ decays. Furthermore the measurement of the branching fraction of $B^{0}\to \rz\rz$ decays and fraction of longitudinal polarization in this decay is presented. We use the results to constrain $\phi_2$ with isospin analyses in the $B\to \pi\pi$ and $B\to\rho\rho$ systems. 
\end{Abstract}
\vfill
\begin{Presented}
CKM2014, the 8th International Workshop on the CKM Unitarity Triangle\\
Vienna, Austria
\end{Presented}
\vfill
%\end{titlepage}
\def\thefootnote{\fnsymbol{footnote}}
\setcounter{footnote}{0}

\section{Introduction}
One major precision test of the Standard Model (SM) is to validate the Cabibbo-Kobayashi-Maskawa (CKM) mechanism for violation of the combined charge-parity ($CP$) symmetry~\cite{C,KM}. This is one of the main purposes of the Belle experiment at KEK which has significantly contributed to proving the CKM scheme and constraining the unitarity triangle for $B$ decays to its current precision. Any deviation from unitarity would be a clear hint for physics beyond the SM. These proceedings give a summary of the experimental status of measurements of the CKM angle \phitwo\ defined from the CKM matrix elements as $\phitwo \equiv \arg(-V_{td}V^{*}_{tb})/(V_{ud}V^{*}_{ub})$.

The CKM angles can be determined by measuring the time-dependent asymmetry between \Bz\ and \Bzb\ decays into a common $CP$ eigenstate~\cite{CP}. In the decay sequence, $\YFS \rightarrow B_{CP}B_{\rm Tag} \rightarrow f_{CP}f_{\rm Tag}$, where one of the $B$ mesons decays into a $CP$ eigenstate  $f_{CP}$ at a time $t_{CP}$ and the other decays into a flavour specific final state $f_{\rm Tag}$ at a time $t_{\rm Tag}$, the time-dependent decay rate is given by
\begin{equation}
  {P}(\Delta t, q) = \frac{e^{-|\Dt|/\tau_{B^0}}}{4\tau_{B^0}} \bigg[ 1+q(\Acp\cos\Delta m_d \Dt + \Scp\sin\Delta m_d \Dt) \bigg],
\label{eq1}
\end{equation}
where $\Dt \equiv t_{CP}- t_{\rm Tag}$ is the lifetime difference between the two $B$ mesons, $\Delta m_d$ is the mass difference between the mass eigenstates  $B_{H}$ and $B_{L}$ and $q = +1 (-1)$ for $B_{\rm Tag} = \Bz (\Bzb)$. The $CP$ asymmetry is given by 
\begin{equation}
\frac{N(\bar{B}\to f_{CP}) - N(B\to f_{CP})}{N(\bar{B}\to f_{CP}) + N(B\to f_{CP})},
\label{eq_asym}
\end{equation}
where $ N(B(\bar{B})\to f_{CP})$ is the number of events of a $B(\bar{B})$ decaying to $f_{CP}$, the asymmetry can be time-dependent.
 The parameters \Acp\ and \Scp\ describe direct and mixing-induced $CP$ violation, respectively~\footnote{There exists an alternate notation where $C_{CP} = -\Acp$.}.  All measurements presented here are based on Belle's final data set of $772 \times 10^{6}$ $B\bar{B}$ pairs.

\section{The CKM Angle \phitwo}
\begin{figure}[htb]
  \centering
  \includegraphics[width=.6\textwidth]{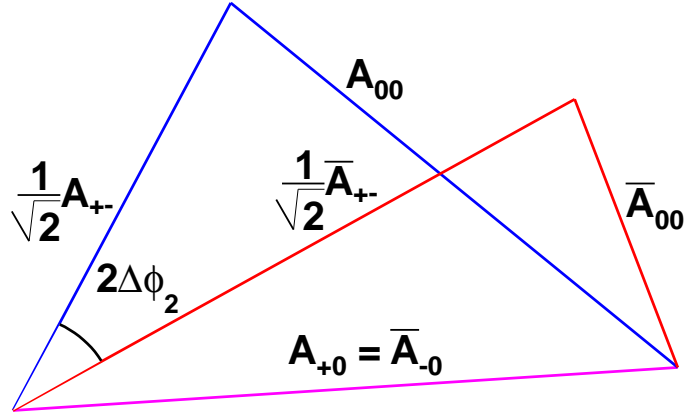}
  \caption{A sketch of the isospin triangle.}
  \label{fig_iso}
\end{figure}

Decays proceeding via $b \rightarrow u \bar{u} d$ quark transitions such as $\Bz \rightarrow \pi\pi$, $\rho\pi$, $\rho\rho$ and $\aone\pi$, are directly sensitive to \phitwo. At tree level we expect $\Acp=0$ and $\Scp=\sin2\phitwo$. Possible penguin contributions can give rise of direct $CP$ violation, $\Acp\neq 0$ and also pollute the measurement of \phitwo, $\Scp=\sqrt{1-\Acp^{2}}\sin(2\phitwo^{eff})$ where the observed $\phitwo^{eff} \equiv \phitwo - \Delta \phitwo$ is shifted by $\Delta \phi_2$ due to different weak and strong phases from additional non-leading contributions.
\begin{figure}[htb]
  \centering
  \includegraphics[width=.5\textwidth]{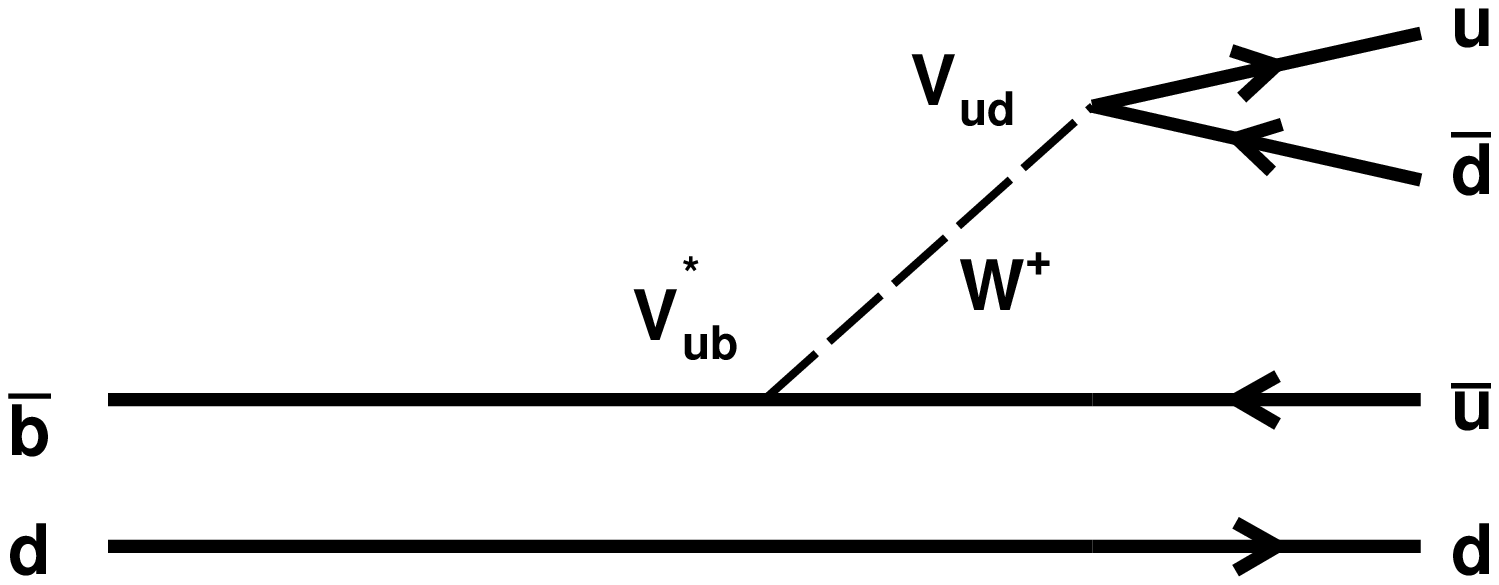}
  \includegraphics[width=.35\textwidth]{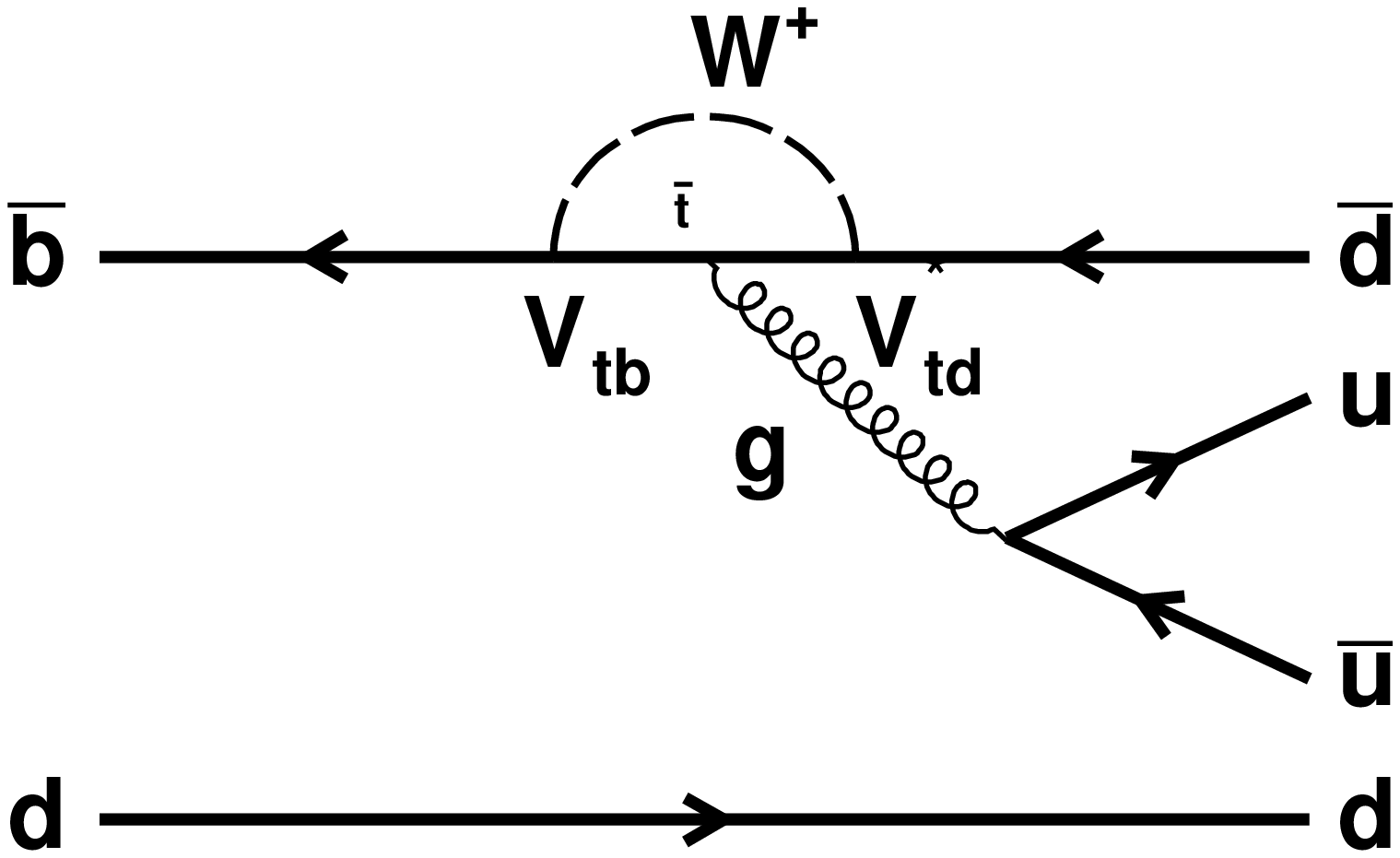}
 \put(-380,1){\scriptsize a)}
 \put(-160,1){\scriptsize b)}\\
  \caption{a) leading order and b) penguin feynman diagrams for  $b \rightarrow u \bar{u} d$ transitions.}
  \label{fig_pipi_feyn}
\end{figure}
Despite this, it is possible to determine $\Delta \phitwo$ in $\Bz \rightarrow h^{+} h^{-}$ with an $SU(2)$ isospin analysis by considering the set of three $B \rightarrow hh$ decays where the $hh$s are either two pions or two longitudinally polarized $\rho s$, related via isospin symmetry~\cite{iso}. The $B \rightarrow hh$ amplitudes obey the triangle relations,
\begin{equation}
  A_{+0} = \frac{1}{\sqrt{2}}A_{+-} + A_{00}, \;\;\;\; \bar{A}_{-0} = \frac{1}{\sqrt{2}}\bar{A}_{+-} + \bar{A}_{00}.
  \label{eq_iso}
\end{equation}
Isospin arguments demonstrate that $\Bp \rightarrow h^{+} h^{0}$ is a pure first-order mode in the limit of neglecting electroweak penguins, thus these triangles share the same base, $A_{+0}=\bar{A}_{-0}$, see Fig.~\ref{fig_iso} for an illustration. $\Delta \phitwo$ can then be determined from the difference between the two triangles. This method has an inherent 8-fold discrete ambiguity in the determination of \phitwo.
%All results presented in this section are preliminary.

\section{The Decay $B^0\to\pip\pim$}
\begin{figure}[htb]
  \centering
 \includegraphics[width=.49\textwidth]{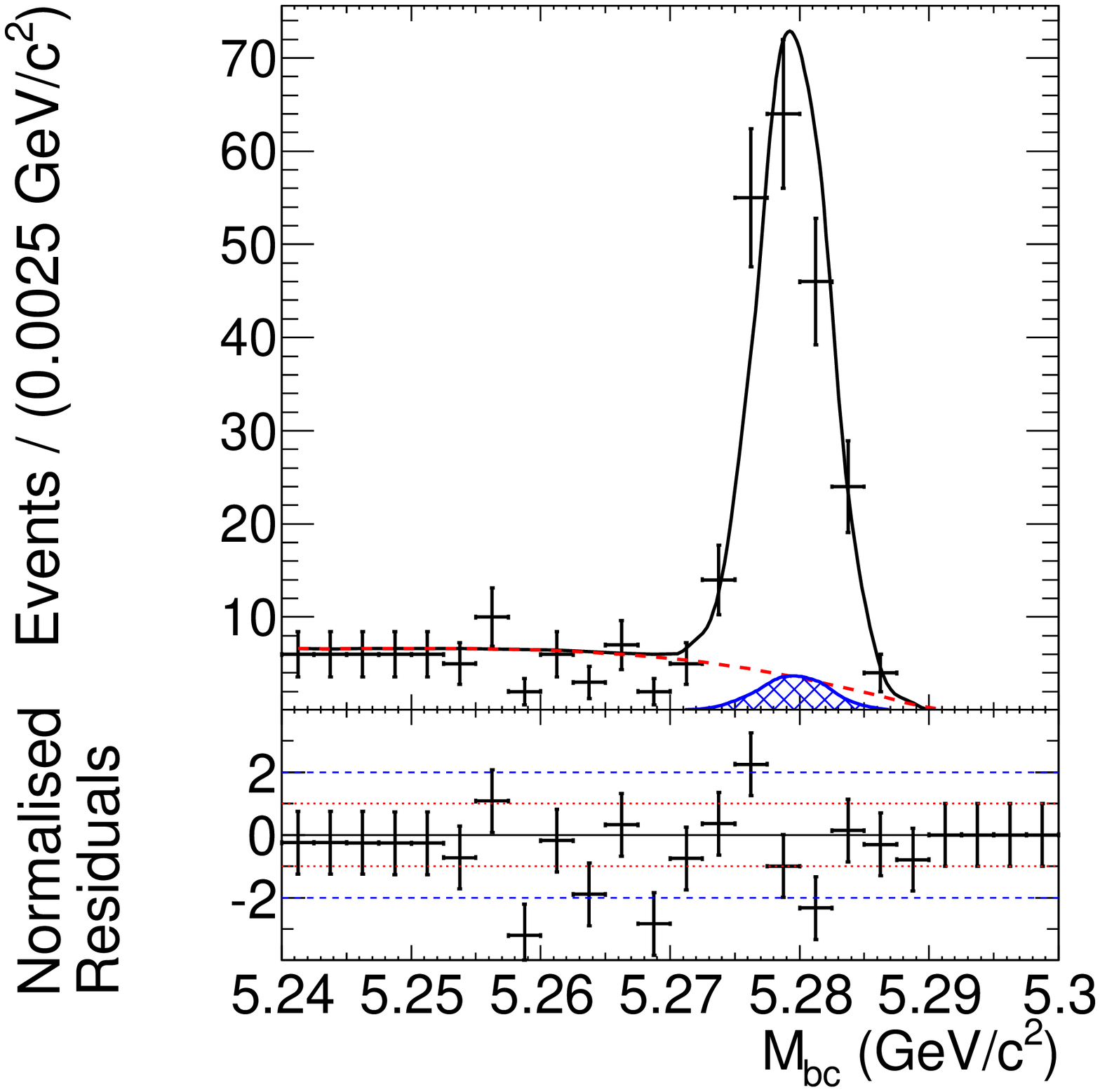}
  \includegraphics[width=.49\textwidth]{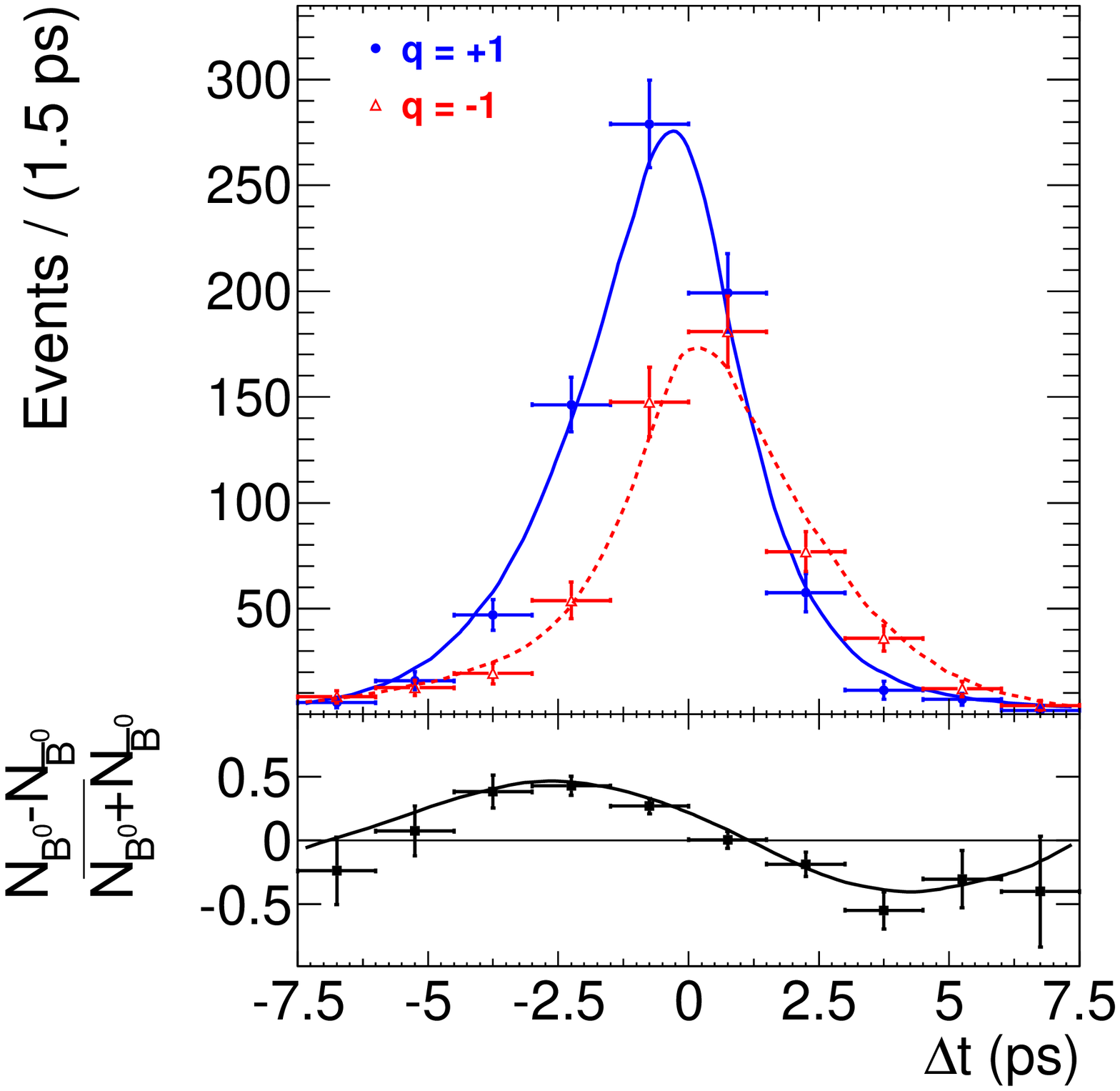}
  \put(-410,20){\scriptsize a)}
 \put(-210,20){\scriptsize b)}\\
  \caption{a) \Mbc\ distribution and b) \Dt\ distribution for each flavour tag and the fit result on top. b) also shows the resulting $CP$ asymmetry for \pippim. Mixing-induced $CP$ violation can be clearly seen in the asymmetry plots and the height difference in the \Dt\ projection indicates direct $CP$ violation. }
  \label{fig_pipi_tcpv}
\end{figure}

\begin{figure}[htb]
  \centering
  \includegraphics[width=.49\textwidth]{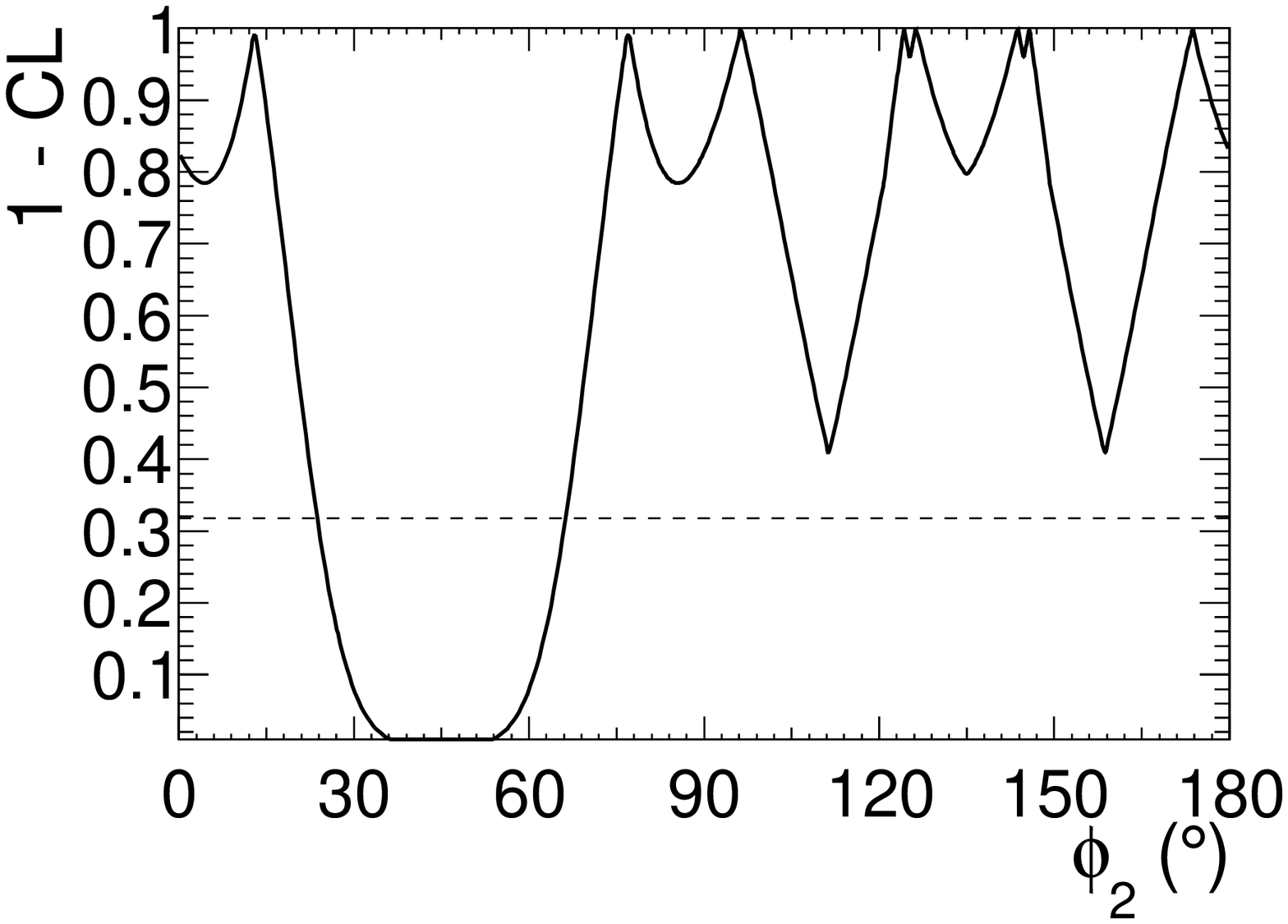}
 \includegraphics[width=.49\textwidth]{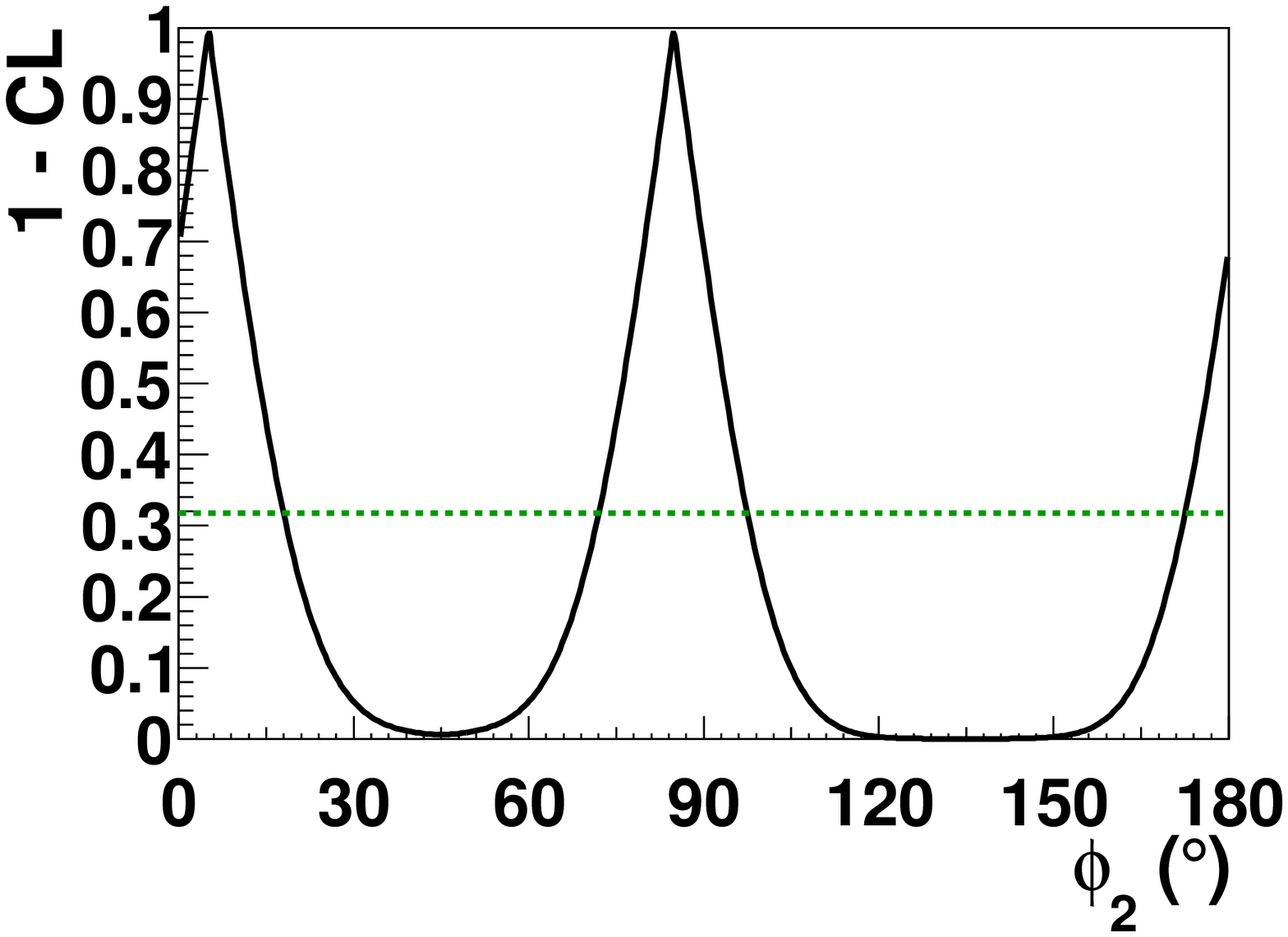}
\put(-400,5){ a)}
 \put(-180,5){ b)}\\
  \caption{Scans of \phitwo\ from an isospin analysis in the a) $B\to \pi\pi$ and b) $B\to\rho\rho$ system, the dashed line corresponds to the one $\sigma$ level.}
  \label{fig_phi2}
\end{figure}

A seven-dimensional extended maximum likelihood fit to the difference of the energy of the reconstructed $B$ meson to the beam-energy (\De), the beam-energy constraint $B$ mass (\Mbc), two likelihoods for each charged track to be a Kaon assuming a $\pi^\pm$ mass hypothesis ($L_{K/pi}^{\pm}$), a event shape-dependent fisher discriminant (${\cal F}_{B\bar{B}/q\bar{q}}$) which separates $B\bar{B}$ (spherical) from $q\bar{q}$ (two-jet like) events, the life-time difference of $B_{CP}$ and $B_{\rm tag}$ (\Dt) and the flavor of $B_{\rm tag}$ ($q$), with $q=+1(-1)$ for $B_{\rm tag}=B^0(\bar{B^0})$ was performed to simultaniously obtain the yields and $CP$ asymmetries of $B\to h^+h^-$ ($h=K,\pi$) decays~\cite{pipi_Belle}. The measurement of the $CP$ parameters in $B^0\to\pip\pim$ decays yield ${\cal S}_{CP} = -0.636 \pm 0.082\;(\rm stat) \pm 0.027\;(\rm syst)$ and ${\cal A}_{CP} = 0.328 \pm 0.061\;(\rm stat) \pm 0.027\;(\rm syst)$. The \Mbc\ and $\Delta t$ distributions are shown in Fig.~\ref{fig_pipi_tcpv}. Belle excludes the range $\phi_2 \notin [23.8^{\circ}, 66.8^{\circ}]$ at the $1 \sigma$ level by performing an isospin analysis to remove the penguin contribution, see Fig.~\ref{fig_phi2}~a). The amount of direct $CP$ violation was found to be smaller compared to the previous measurement at Belle~\cite{pipi_Belle_old}. The previous result was confirmed with the previous data set of $535\times 10^{6}$ $B\bar{B}$ pairs. While their was already good agreement in previous measurements of ${\cal S}_{CP}$, the drop in ${\cal A}_{CP}$ leads to a in better agreement with other experiments~\cite{pipi_BaBar,pipi_LHCb}. This is currently the most precise measurement of $CP$ violation in $B^0\to\pip\pim$ decays.

\section{The Decay $B^0\to\piz\piz$}
We only report on this mode briefly as it is discussed in detail in another presentation (M.Seviour, WG $V$). This decay is reconstructed from four $\gamma$s making this measurement experimentally quite challenging. A fit to \De, \Mbc\ and ${\cal F}_{B\bar{B}/q\bar{q}}$ is performed and a preliminary branching fraction of ${\cal B}(B\to\piz\piz) = (0.9\pm 0.12\;{\rm stat}\pm 0.10\;{\rm syst})$ is obtained. It is planned to supersede this measurement with one including a measurement of direct $CP$ violation in this mode. For next generation $B$ factories photon conversion makes vertexing possible and allows to perform a measurement of the time-dependent $CP$ asymetries.

\section{The Decay \rhozrhoz}
\begin{figure}[htb]
  \centering
  \includegraphics[width=.45\textwidth]{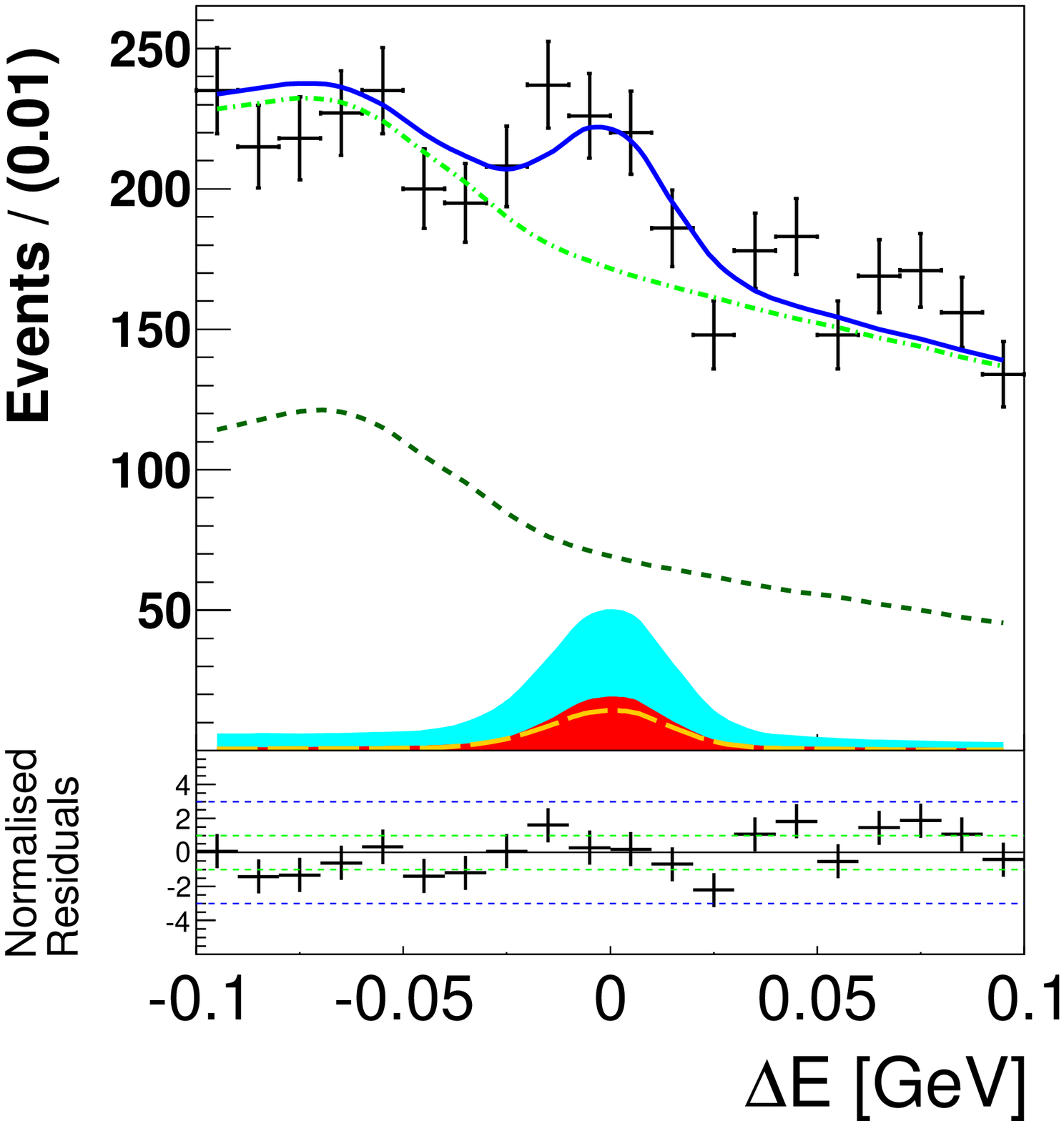}
  \includegraphics[width=.45\textwidth]{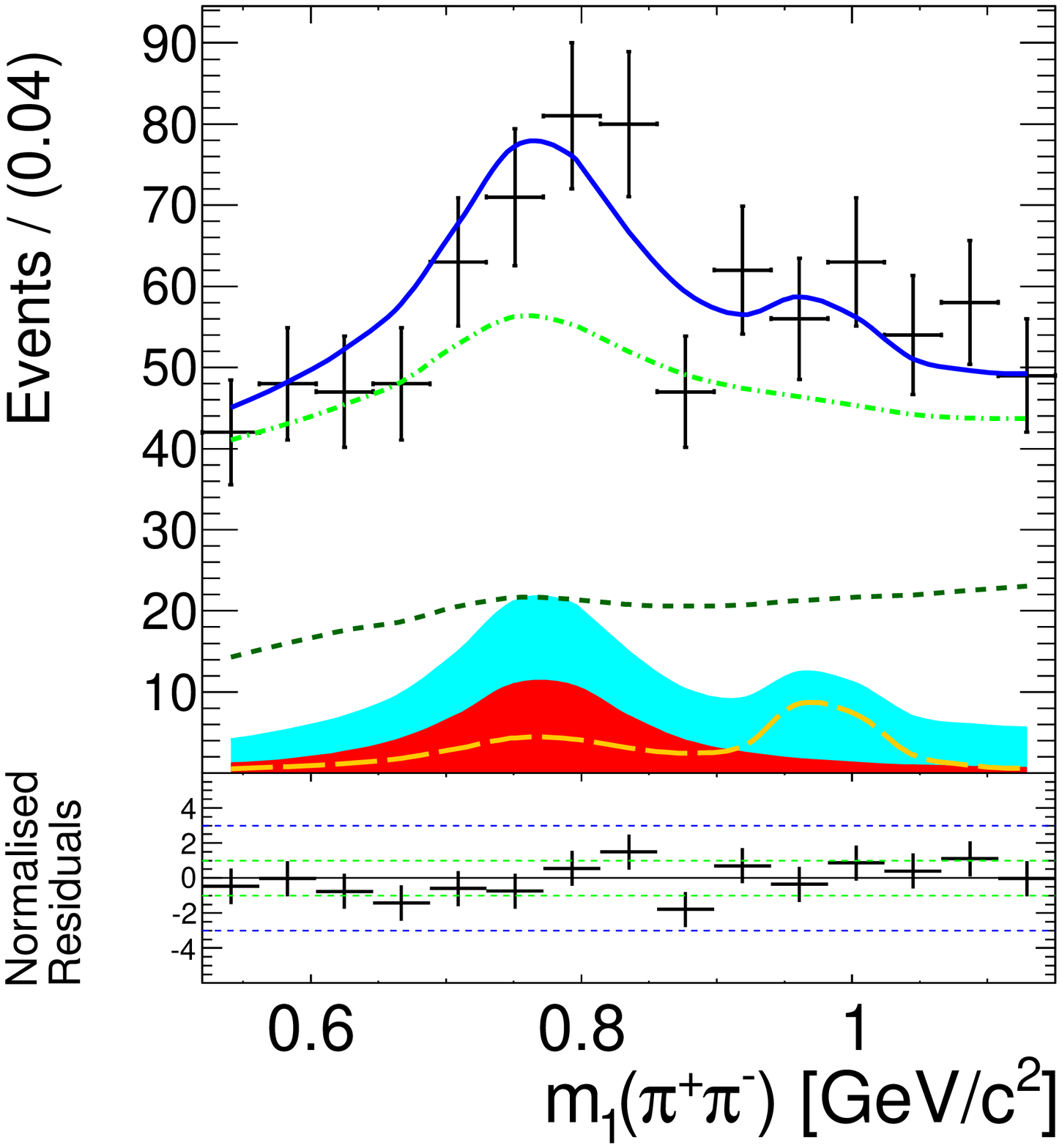}
\put(-400,5){ a)}
 \put(-180,5){ b)}\\
  \includegraphics[width=.45\textwidth]{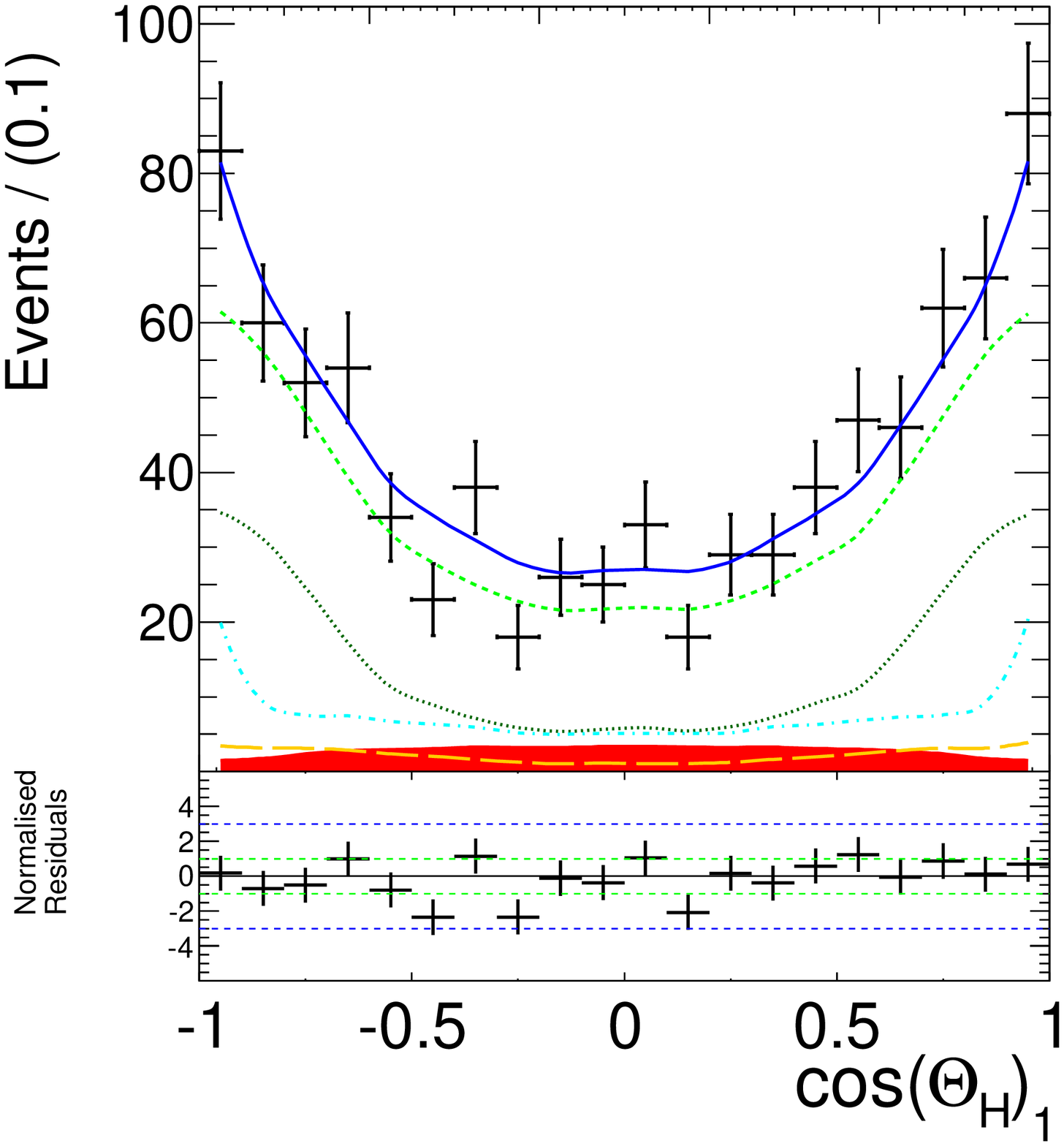}
\includegraphics[width=.45\textwidth]{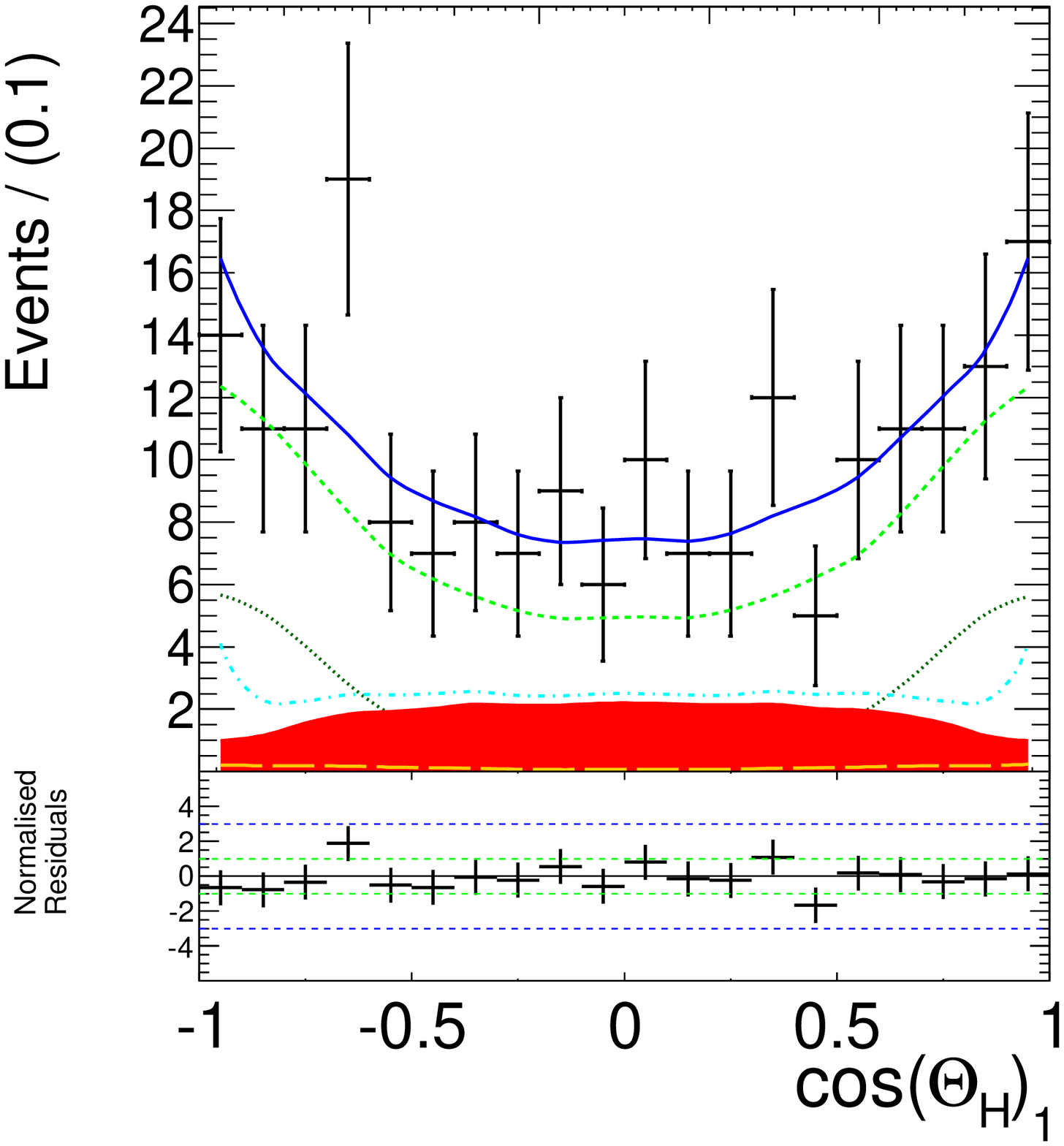}
\put(-400,5){c)}
 \put(-180,5){ d)}\\
  \caption{ Signal enhanced distributions of a) $\Delta E$, b) $m(\pip\pim)$ and c,d) $\cos\Theta_H$ with the fit result on top. d) shows a distribution where the contributions from the remaining four-pion final states are reduced by a tight cut on the $m(\pip\pim)$-$m(\pip\pim)$ distribution, demonstrating the necessity of a transversaly polarized component. The shaded red area and the long dashed orange histogram are the $B^0\to\rho^0\rho^0$ and $f_0\rho^0$ contributions, respectively. Furthermore, all four pion final states are shown in dashed cyan, the entire ($B\bar{B}$) background in dashed green (dark green) and the full PDF in blue.}
  \label{p_r0r0}
\end{figure}

 Having a decay into two vector particles, an angular analysis has to be performed in order to separate the pure $CP$-even states with longitudinal polarization from the transverse polarized mixture of $CP$-even and odd states for the isopsin analysis.
The presence of multiple, largely unknown backgrounds with the same four charged pions final state as $B^0\to\rho^0\rho^0$ make this decay quite difficult to isolate and interference between the various $4\pi$ modes needs to be considered. Similar as for $B^{0}\to\rho^{+}\rho^{-}$ decays~\cite{rhoprhom_BaBar, rhoprhom_Belle1, rhoprhom_Belle2}, the decay \rhorho\ is naively expected to be polarized dominantly longitudinally. However, color-suppressed $B$ decays into light vectors are especially difficult to predict as the transverse polarized amplitude is not calculable~\cite{BVV_theo} .
Besides updating to the full data set, two helicity angles $\cos\Theta_H^{1,2}$, one for each $\rho^{0}$, are added to the fit. The angles, defined in the helicity basis, are powerful in separating the different backgrounds and allow one to measure the fraction $f_L$ of longitudinal polarization. Belle obtains ${\cal B}(B^0\to\rho^0\rho^0) = (1.02 \pm0.30\;(\rm stat) \pm 0.15\;(\rm syst)) \times 10^{-6}$ with a fraction of longitudinal polarization, $f_L = 0.21^{+0.18}_{-0.22}\;(\rm stat)\pm 0.15\;(\rm syst)$~\cite{belle_r0r0}. Signal enhanced distributions onto $\Delta E$, $m(\pip\pim)$ and one of the helicty angles, each with the fit result on top, are shown in Fig.~\ref{p_r0r0}.
 Since this mode is found to decay dominantly into transversely polarized $\rho^0$s, a measurement of the $CP$ asymmetries has not been performed. However, the size of the amplitude of the decays into longitudinally polarized $\rho^0$s from this measurement has been used in an isospin analysis together with other Belle measurements~\cite{rhoprhom_Belle1, rhoprhom_Belle2, rpr0_Belle} (longitudinal polarization only). The resulting constraint most consistent with other measurements of the CKM triangle is $\phitwo = (84.9 \pm 13.5)^{\circ}$. The relatively small amplitude of $B^0\to\rho^0\rho^0$ makes the isospin analysis in the $B\to\rho\rho$ less ambiguous. Fig.~\ref{fig_phi2}~b) shows \phitwo\ scan from the isospin analysis.
Comparing these results with the ones obtained by BaBar, we find good agreement in the branching fraction of $\rhozrhoz$ decays, while there is a $2.1\sigma$ discrepancy in the fraction of longitudinal polarization. BaBar finds $f_L = 0.75^{+0.11}_{-0.14}\;(\rm stat)\pm 0.04\;(\rm syst)$~\cite{r0r0Babar}. This measurement is still statistically (and systematically) limited, hence further studies with more statistics would be very interesting and hopefully will solve this tensions. 

\section{Summary}
We have presented recent measurements from Belle sensitive to the CKM phase \phitwo\ using the full data set.
Measurements of the $CP$ asymmetries in $B \rightarrow \pip \pim$ and the branching fraction of $B\to\rho^0\rho^0$, together with fraction of longitudinal polarized $\rho^0$s in this decay were presented. The data are used to constrain \phitwo\ with an $SU(2)$ isospin analysis for each isospin triplet. The current world averages of \phitwo\ as computed by the CKMfitter~\cite{CKMfitter} (including the results presented) and UTfit~\cite{UTfit} collaborations are $\phitwo = (88.5^{+4.7}_{-4.4})^{\circ}$ and $\phitwo = (89.1 \pm 3.0 )^{\circ}$, respectively. Furthermore we presented a preliminary update of the branching fraction of $B^0\to\piz\piz$ decays. With BelleII being built and LHCb operating, the next generation of $B$ physics experiments are expected to further reduce the uncertainty of the CKM observables, e.g. the uncertainty of \phitwo\ is expected to be reduced to $\sim 1^{\circ}$~\cite{Belle2}.

\end{document}